\documentclass[aps,prb,twocolumn,superscriptaddress]{revtex4-2}

\usepackage{graphicx}
\usepackage{dcolumn}
\usepackage{bm}
\usepackage{amsmath}
\usepackage{amssymb}
\usepackage{cases}

\usepackage{color}

\newcommand{\He}{$^4$He}
\newcommand{\et}{$\it{et \textrm{ } al. }$ }
\newcommand{\ie}{$\it{i.e.}$}

\newcommand{\B}[1]{\bm{#1}}

\begin{document}


\title{Coupled dynamics of quantized vortices and normal fluid in superfluid \He{} based on lattice Boltzmann method}



\author{Sosuke Inui}
\affiliation{Department of Physics, Osaka City University, 3-3-138 Sugimoto, 558-8585 Osaka, Japan}
\author{Makoto Tsubota}
\affiliation{Department of Physics \& Nambu Yoichiro Institute of Theoretical and Experimental Physics (NITEP) \& The OCU Advanced Research Institute for Natural Science and Technology (OCARINA), Osaka City University, 3-3-138 Sugimoto, 558-8585 Osaka, Japan}

\date{\today}

\begin{abstract}
We investigate the coupled dynamics of quantized vortices and normal fluid in superfluid \He{} at finite temperatures using a numerical approach based on the vortex filament model (VFM) and lattice Boltzmann method (LBM).
The LBM allows us to simulate a fluid flow with only local operations, \ie{}, rather than solving the Navier--Stoles (NS) equations directly; a fluid flow is considered a convection of mesoscopic particles between sites on a lattice grid.
Although the two-fluid nature of He II makes its flow complex, the particle-like treatment of the normal fluid in the LBM significantly reduces the complexity.
We confirm, by comparing to results obtained with direct NS simulations, that the proposed numerical approach reproduces characteristic flow structures.
We also demonstrate that the proposed computational approach is suitable for a thermal counterflow simulation with a solid boundary to elucidate a thermal boundary layer near a heater in a closed channel.
\end{abstract}

\pacs{xxxx}

\maketitle


\section{Introduction}
%
Since the discovery of superfluidity in the 1930’s, its fascinating properties such as vanishing shear viscosity and extremely high thermal conductivity have attracted the attention of a considerable number of physicists and engineers for various purposes.
For physicists in the field of fluid dynamics, a superfluid is essentially an ideal fluid where the flow structures are significantly simplified; it is hoped that it can have an important role in solving and understanding the problems of turbulence.
The properties of superfluid \He{} are also appreciated in engineering fields that require extremely high-magnetic fields.
 Because the thermal conductivity of superfluid \He{} is several million times higher than that of normal liquid helium above the transition temperature $T_\lambda = 2.17$ K, it transfers heat efficiently without boiling, which makes the superfluid \He{} an excellent coolant that can maintain constantly high superconducting magnet performance \cite{VanSciver, Sciacca2015}.
%

The idea of a two-fluid nature introduced by Titza \cite{Tisza38} and Landau \cite{Landau41} explains the high thermal conductivity of superfluid \He{} as a consequence of the internal convection of the superfluid and normal fluid components.
The entire fluid at $0$ K is an inviscid superfluid, where macroscopic quantum effects govern the system; at finite temperatures, the system is a mixture of inviscid superfluid of density $\rho_s$ and viscous normal fluid of density $\rho_n$.
The normal fluid is composed of thermal excitations, \ie{}, phonons and rotons, and carries nonzero entropy \cite{Tilley90, Donnelly91}.
From a simple thermodynamical argument, the relation between the heat flux $\B{q}$ and collective motion of such thermal excitations, or spatially averaged normal fluid velocity $\B{v}_n$, can be derived as follows:
\begin{equation} \label{eq: q and v_n}
\B{q} = \rho \B{v}_n \sigma T,
\end{equation}
where $\rho \equiv \rho_s + \rho_n$ is the total density of the fluid, $\sigma$ is the temperature-dependent specific entropy of the normal fluid, and $T$ is the temperature of the system.
This relationship has been verified experimentally in thermal counterflow experiments for decades \cite{Gorter49,Vinen57_1,Vinen57_2,Vinen57_3,Vinen57_4,Brewer,Chase,Childers,Tough82}.
The concept of the experiment is as follows. We consider a long closed pipe of cross-sectional area $A$ filled with superfluid \He{} at temperature $T$. 
Then, by turning on a heater of heating power $W$ placed near one end of the pipe, we can excite a normal fluid that travels away from the heater, while the superfluid is driven toward the heater to cancel the total mass flux.
Based on Eq. \eqref{eq: q and v_n}, the relative velocity $v_{ns} = |\B{v}_n - \B{v}_s|$ in a steady state is 
\begin{equation} \label{eq: v_ns and W}
v_{ns} = \frac{1}{\rho_s \sigma T} \frac{W}{A}.
\end{equation}
However, for heating powers greater than some critical value $W_c$, the relative velocity $v_{ns}$ generated becomes turbulent and the straightforward argument is no longer valid.
%

The turbulent state of a superfluid component is often identified with a tangled structure of quantized vortices.
More precisely, several turbulent regimes exist in a channel thermal counterflow, such as the $T$-1 and $T$-2 turbulent regimes \cite{Tough82}.
In the $T$-1 regime, only the superfluid component is considered to be turbulent, whereas the normal-fluid component remains laminar.
In the $T$-2 regime, conversely, both components are turbulent.
The indispensable ingredient of superfluid turbulence, namely a quantized vortex, is a filamentary topological defect around which the circulation $\Gamma = \oint_\mathcal{L} d\B{l} \cdot \B{v}_s$ along a path $\mathcal{L}$ is quantized to be integer-multiples of $\kappa = h/m$, where $h$ is Planck's constant and $m$ is the mass of a \He{} atom.
In superfluid \He{}, the circulation quantum number of all the vortices is assumed to be $n=1$ as they are energetically preferable to $n=2$ vortices.
The core radius of such vortices is known to be of the order of an $\AA$ and is negligible compared to the typical experimental setup size scale (a few mm to cm).
Therefore, we apply the vortex filament model (VFM) to model the dynamics of quantized vortices.
The VFM is a powerful computational tool that significantly reduces the complexity of the dynamics, yet allows the reproduction of quantum turbulent states under different settings.
Although such preceding computational results have enabled us to understand the steady quantum turbulence quantitatively, in the majority of the early approaches, quantized vortices were considered as completely independent objects, \ie{}, the normal fluid velocity profile was prescribed and the vortices were influenced by the normal fluid through mutual friction, whereas the vortices did not influence the dynamics of the normal fluid.
Several recent studies have addressed this issue by solving the Hall--Vinen--Bekarevich--Khalatnikov (HVBK) equations for normal fluid component \cite{Biferale19, Bertolaccini17, Yui18,Kobayashi19, Yui20,Galantucci20,Galantucci21}.
The HVBK equations are essentially the Navier--Stokes (NS) equations with a forcing term ascribed to the mutual friction between the normal and superfluid components due to the motions of the quantized vortices \cite{Donnelly91}.
%

In this study, we suggest a different approach to address the coupled dynamics of normal fluid and superfluid flow from the preceding studies, \ie{}, we apply the lattice Boltzmann method (LBM) for the normal fluid and VFM for the superfluid dynamics.
The LBM is frequently called a ``mesoscopic'' model because it considers fluid motion as the advection of mesoscopic fluid particles from a lattice site to another neighboring site and the collisions between them.
The LBM is widely used in computational fluid dynamics (CFD) and several of its variations have been developed to reproduce classical fluid flows under different conditions for decades.
In Ref. \cite{Bertolaccini17}, Bertolaccini \et{} apply the LBM to both fluids (super and normal) and investigate the coupled dynamics under a thermal counterflow in a two dimensional channel. 
Unlike their approach, we apply a three-dimensional LBM for only the normal-fluid component; we follow the dynamics of a superfluid based on the VFM such that the local superfluid flow information is not lost by the averaging processes.
%

Detailed explanations of the numerical approaches are presented in Sec. \ref{sec: NS}.
In Sec. \ref{sec: RES}, we present several fundamental motions of vortices coupled with the normal-fluid component and we discuss the numerical results that reproduce those of several recent studies.
In particular, we consider the dynamics of a single vortex ring, vortex-vortex reconnection events, and a vortex tangle development under a thermal counterflow in a periodic channel.
One of the most noticeable advantages of the LBM approach is its flexibility in the treatment of boundaries/surfaces.
 One of the possibilities is a spherical heater that is immersed in a fluid and creates a steady thermal counterflow.
As discussed in Ref. \cite{Inui20_1}, a numerical study with such a boundary condition could allow us to obtain a quantitative understanding on the ``micro big-bang'' experiment in Grenoble \cite{Bauerle96,Bunkov14}.
To examine the validity of the proposed numerical approach near the surface, in Sec. \ref{sec: DIS}, we consider a simple (but not trivial) boundary condition, \ie{}, planar heaters with solid surfaces in a closed channel (see Fig. \ref{fig: Wall_density_anim}), rather than more involved boundaries.
In Sec. \ref{sec: SUM} we summarize what we have presented in this article.

\section{Numerical Approach} \label{sec: NS}
%
\subsection{Vortex filament model (VFM)}
A quantized vortex in superfluid \He{} at $0$ K travels with the background superfluid velocity field according to Helmholtz's law \cite{Helmholtz1868}.
However, at a finite temperature, thermal excitations that compose the normal fluid hinder the vortex from traveling accordingly, which is ascribed to the origin of the mutual friction between the two fluids.
Consider a set of position vectors $\B{s} = \B{s}(\xi,t)$ that represent a vortex with an infinitesimal core size at a time $t$.
Its spatial positions are parameterized by its arc length $\xi$; $\xi$ is discretized into segments of lengths between $\Delta \xi_\textrm{max}$ and $\Delta \xi_\textrm{min}$.
The superfluid velocity $\B{v}_{s}$ induced by the vortex at some position $\B{r}$ is given by the Biot--Savart integral along path $\mathcal{L}$ that corresponds to the vortex position vectors.
\begin{equation} \label{eq: Biot--Savart}
\B{v}_{s}(\B{r}) = \frac{\kappa}{4 \pi} \int_\mathcal{L} d\xi \frac{(\B{s}(\xi) - \B{r}) \times \B{s}^\prime}{| \B{s}(\xi) - \B{r} |^3} + \B{v}_{s,b} + \B{v}_{s,a},
\end{equation}
where the prime symbol represents a derivative with respect to $\xi$.
$\B{v}_{s,b}$ represents the superfluid velocity field induced by a given boundary condition and $\B{v}_{s,a}$ represents the externally applied velocity field.
The integral is performed segment-wise in the numerical simulation.
Because Eq. \eqref{eq: Biot--Savart} is divergent around a point where $\B{s} \rightarrow \B{r}$, that portion of the integral is calculated separately.
The separation of the divergent local term is often referred to as the local induction approximation (LIA); this approach was first applied by Schwarz for numerical simulations of superfluidity \cite{Schwarz85,Schwarz88}.
In recent studies \cite{Adachi10,Yui20}, including this work, the full Biot--Savart integral, \ie{}, the LIA term $+$ nonlocal term, is calculated to consider the nonlocal behavior of the vortices.
The equation of motion for a vortex segment at finite temperatures is found as
\begin{equation} \label{eq: E of M}
\frac{d \B{s} }{d t} = \B{v}_s + \alpha \B{s}^\prime \times (\B{v}_{n} - \B{v}_{s})- \alpha^\prime  \B{s}^\prime \times  \left[ \B{s}^\prime \times  (\B{v}_{n} - \B{v}_{s})  \right],
\end{equation}
where $\alpha$ and $\alpha^\prime$ are the temperature dependent mutual friction coefficients.
Thus, the time evolution of the vortices can be obtained by solving this integro-differential equation with an initial condition.

The idealized vortex filaments, governed by Biot--Savart law, do not $\it{naturally}$ reproduce any vortex-reconnection event by themselves. 
 However, it is widely observed that such reconnection phenomena do take place in superfluid \He{} experimentally \cite{Bewley06,Bewley08,Fonda19} and in numerical simulations based on Gross--Pitaevskii equation \cite{Koplik93,Zuccher12}.  
 We, therefore, algorithmically exchange the legs of vortices when two vortices approach each other within the spatial resolution $\Delta \xi$.  
 Although there is some arbitrariness in choice of reconnection algorithms, VFM is known to be robust to the choice of algorithm and the statistical properties of the vortex tangle do not depend on the choice \cite{Baggaley12}.

\subsection{Lattice Boltzmann method (LBM)}
The first numerical simulation of a flow of superfluid \He{} based on the LBM was performed by Bertolaccini \et{} \cite{Bertolaccini17}.
The LBM considers the flow as a convection of particles with certain discretized momenta.
In the LBM approach, we follow the change in the particle populations in the discretized phase space.
After particle collisions occur at the lattice sites, we calculate a particle population in each momentum state, instead of directly addressing the continuum equations.
Conventional continuum equations can be recovered by averaging over the discretized momentum space.
This implies that one of the main difficulties in solving the NS equation, \ie{}, addressing the convection term $(\B{v}\cdot \B{\nabla} ) \B{v}$, is essentially avoided.
%

There are several widely appreciated variations of discretization approaches for simulating a three-dimensional flow stably \cite{Kruger}.
In this study, we adopt the D3Q19 lattice, where the phase space is defined as a three-dimensional lattice $\B{x} \equiv (x,y,z) \in \mathbb{Z}^3$ with $19$ discrete momenta at each lattice point.
The particle population in the $i^{\textrm{th}}$ momentum state at time $t$ is denoted as $f_i(\B{x},t)$, and the total particle number at a lattice point, or the density $\rho(\B{x},t)$, is thus
\begin{equation}
\rho(\B{x},t) = \sum_{i = 1}^{19} f_i(\B{x},t).
\end{equation}
 Similarly, the local velocity is
 \begin{equation}
\B{u}(\B{x},t) =  \frac{1}{\rho}\sum_{i = 1}^{19}  \B{c}_i  f_i(\B{x},t),
\end{equation}
where $\B{c}_i$ is the velocity of the particles in the $i^{\textrm{th}}$ momentum state. 
The higher moments of $\B{c}_i$'s can be also calculated in a similar manner.
%

\begin{figure}[b!]
	\includegraphics [width=0.7\columnwidth]{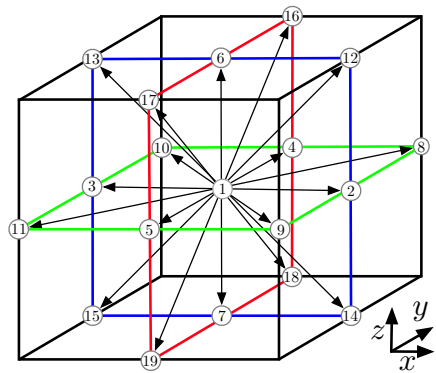}
	\caption{Schematics of momentum states in LBM D3Q19 approach. The arrows correspond to each $\B{c}_i$ in Table \ref{tab: coeffs}.}
	\label{fig: D3Q19}
\end{figure}

The time evolution of the fluid is achieved by the following three steps in the LBM approach. 
Step 1. Calculate the momentum state-wise distribution function $f_i(\B{x})$ and an equilibrium distribution $f_i^\textrm{eq}(\B{x})$ evaluated from the averaged macroscopic quantities as
\begin{equation}
f^{\textrm{eq}}_i(\boldsymbol{x}) = w_i \rho \left( 1 + \frac{\boldsymbol{u}\cdot \boldsymbol{c}_i}{c_s^2} +  \frac{ ( \boldsymbol{u}\cdot \boldsymbol{c}_i ) ^2 }{2c_s^4} -  \frac{\boldsymbol{u}\cdot\boldsymbol{u}}{2c_s^2} \right),
\end{equation}
where $c_s = \frac{1}{\sqrt{3}} \frac{\Delta x}{\Delta t}$ is the characteristic velocity, and $w_i$ and $\B{c}_i$ are the model-dependent coefficients, which are listed in Table \ref{tab: coeffs} and are schematically displayed in Fig. \ref{fig: D3Q19}.
$\Delta x$ and $\Delta t$ are the spatial and temporal resolutions of the LBM, respectively; both are set to unity in this study.
\begin{table}[b!]
\caption{\label{tab: coeffs} Table of coefficients. Lattice units are assumed to be unity; hence, $|\B{c}_i| = \Delta x /\Delta t = 1$ throughout this study.}
\begin{ruledtabular}
\begin{tabular}{ l l l} 
Number &\textrm{Velocity $\B{c}_i$} & \textrm{Weight $w_i$} \\
\colrule \colrule
$i = 1$ & $\B{c}_i = (0,0,0)$ & $w_i = 1/3$ \\
\colrule
$\quad$ & $\B{c}_i = (\pm 1,0,0)$ & $    \quad   $ \\
$i = 2, \dots, 7$ & $\B{c}_i = (0,\pm 1,0)$ & $w_i = 1/18$ \\
$\quad$ & $\B{c}_i = (0,0,\pm 1,)$ & $      \quad   $ \\
\colrule
$\quad$           & $\B{c}_i = (\pm 1,\pm 1,0)$ & $    \quad   $ \\
$i = 8, \dots,19$ & $\B{c}_i = (\pm 1,0,\pm 1)$ & $w_i = 1/36$ \\
$\quad$           & $\B{c}_i = (0,\pm 1,\pm 1)$ & $      \quad   $ \\
\end{tabular}
\end{ruledtabular}
\end{table}
%
Step 2. Compute collisions between the fluid particles at every lattice site.
In classical Boltzmann dynamics, the collision integral $C[f]$ is involved, which is, in general, a nontrivial multidimensional integral.
In the LBM, however, the collision integral is significantly simplified by employing the distribution functions found in Step 1. 
In this study, we approximate the collision integral such that it becomes linear in $f_i$ as
\begin{equation}\label{eq: BGK operator}
C[f_i] =  - \frac{\Delta t}{\tau} \left( f_i - f^{\textrm{eq}}_i    \right),
\end{equation}
where, $\tau$ is the relaxation time, which essentially determines how quickly $f_i$ converges to the local equilibrium states $f_i^\textrm{eq}$.
The approximated collision integral in Eq.\eqref{eq: BGK operator} is known as the Bhatnagar--Gross--Krook (BGK) collision operator and allows us to reduce the computational cost significantly.
Equation \eqref{eq: BGK operator} implicitly requires that the distribution function $f_i$ at each lattice site in the system be sufficiently close to those at the local equilibria.
Therefore, to simulate $T$-2 turbulent states \cite{Tough82}, where both normal fluid and superfluid are turbulent, this approach may not be suitable.
In this study, we only consider the case where the normal fluid is laminar. 
More precisely, the condition of small Mach number ($|\B{u}|/c_s < 0.05$) is imposed so that Eq. \eqref{eq: BGK operator} is valid throughout the simulations.
Step 3. Stream particles at each lattice site to the neighboring lattice sites based on the distribution function $f_i$.
As defined above, the number of particles in the $i^{\textrm{th}}$ momentum state at location $\B{x}$ at time $t$ is $f_i(\B{x},t)$.
Because these particles travel to the next site at $(\B{x}+\B{c}_i \Delta t)$ after $\Delta t = 1$ in the lattice Boltzmann (LB) time unit, the particle population after $\Delta t$ can be asymptotically expressed as
\begin{equation}
 f_i(\B{x},t) \rightarrow f_i(\B{x}+\B{c}_i \Delta t, t + \Delta t) \quad \textrm{as}\quad \Delta t \rightarrow 1,
\end{equation}
if we ignore the collisions between the particles discussed in Step 2.
%

The coupling between the normal fluid and superfluid flows arises from the mutual friction mediated by the quantized vortices.
The last two terms on the left side of Eq. \eqref{eq: E of M} and the mutual friction $\B{f}$ on a vortex per unit length at $\xi$ are related as follows: \cite{Barenghi83,Schwarz85,Yui20};
\begin{equation}
\frac{\B{f}(\xi)}{\rho_s \kappa} =  \alpha \B{s}^\prime \times (\B{v}_{n} - \B{v}_{s})- \alpha^\prime  \B{s}^\prime \times  \left[ \B{s}^\prime \times  (\B{v}_{n} - \B{v}_{s})  \right],
\end{equation}
and its reaction is on the normal fluid.
The reaction on the normal fluid due to a vortex segment at $\xi$ is interpolated at the $8$ surrounding neighboring sites using a discrete two-point delta function
\begin{equation}
\delta_{2\textrm{pt}}(r) = \begin{cases}
1-|r|  & \quad\text{for $|r|\leq 1$}  \\
0                         & \quad \text{for $|r| > 1$.}
\end{cases}
\end{equation}
After the interpolation, the reaction on the normal fluid is converted into a force density $F_{i}(\B{x})$ such that it acts on the distribution function $f_i$ individually.
Thus, we obtain the expression for the governing equation for the LBM by summarizing the above:
\begin{equation}
f_i(\B{x}+\B{c}_i \Delta t, t + \Delta t) = f_i(\B{x},t) - \frac{1}{\tau} \left(   f_i(\B{x},t) -  f_i^{\textrm{eq}}(\B{x},t)   \right) + F_i \Delta t.
\end{equation}  
However, such a steep, delta function-like forcing term $F_i$ tends to unstabilize the numerical simulations.
In order to stabilize the code, several approaches can be used to handle this problem and calculate the forcing term \cite{Kang10}.
In this study, we adopt the split-forcing approach developed by Cheng and Li \cite{Cheng08,Kang10}.
Their approach involves the force density of the form
\begin{equation}
F_i(\B{x}) = \left(  1-\frac{1}{2 \tau}  \right) w_i  \left(  \frac{\B{e}_i -\B{u}}{ c_s^2} + \frac{\B{c}_i  \cdot \B{u}}{c_s^4} \B{c}_i  \right) \cdot \B{f}.
\end{equation}
The force density obtained is applied to the distribution function $f_i$ twice by splitting it in two pieces to stabilize the code.
Below, we summarize the numerical operations as a set of four equations:
\begin{subequations}
\begin{align}
&f_i^\prime(\B{x}, t) = f_i(\B{x},t) + \frac{\Delta t}{2} F_i,\label{eq: LBM 1}\\
&f_i^{\prime\prime}(\B{x}, t) = f_i^\prime(\B{x}, t) - \frac{1}{\tau}   \left(   f_i(\B{x},t) -  f_i^{\textrm{eq}} (\B{x},t)   \right),\label{eq: LBM 2}\\
&f_i^{\prime\prime\prime}(\B{x}, t) = f_i^{\prime\prime}(\B{x},t) + \frac{\Delta t}{2} F_i,\label{eq: LBM 3}\\
&f_i(\B{x}+\B{c}_i \Delta t, t + \Delta t) = f_i^{\prime\prime\prime}(\B{x},t).\label{eq: LBM 4} 
\end{align}
\end{subequations}
Equation \eqref{eq: LBM 1} corresponds to the first application of the forcing term.
Then, we update the change in the distribution functions due to the collision according to Eq. \eqref{eq: LBM 2} (Step 2).
The other half of the forcing term is applied on the updated distribution functions as in Eq. \eqref{eq: LBM 3}, and finally fluid particles are streamed in Eq. \eqref{eq: LBM 4} (Step 3).

\section{Results} \label{sec: RES}
%
In this section, we present the results using this computational approach:
a vortex ring traveling in a normal fluid at rest (see \ref{subsec: One Ring}), a characteristic flow pattern of the normal fluid during the reconnection events between two quantized vortices (see \ref{subsec: Reconnections}), and a vortex tangle that develops under a thermal counterflow (see \ref{subsec: Vortex Tangle}).
For the results below, the dimensions of the computational space are $L^{\textrm{LB}}_x=64\Delta x $, $L^{\textrm{LB}}_y=32\Delta y $, and $L^{\textrm{LB}}_z=32\Delta z$, and the spatial resolutions for the VFM, $\Delta \xi_\textrm{max}$ and $\Delta \xi_\textrm{min}$, are set to $0.95\Delta x $ and $ 0.70\Delta x $ in the LB unit, respectively.
Whereas the LB calculations are performed with the quantities measured in LB units, the VFM calculations are  performed with those in physical units.
The conversion between these two unit systems is achieved using the sound velocity $c$ and length $L_x (= 2$ mm$)$ of the channel.
The temperature $T$ of the system is set to be constant at $T=1.9$ K, even when the counterflow is activated.

\subsection{One Ring}\label{subsec: One Ring}
Kivotides \et{} \cite{Kivotides00} indicated that a vortex ring travelling in a normal fluid at rest produces a ``triple vortex ring structure'', which involves a superfluid quantized vortex and two normal-fluid vortex rings created inside and outside the quantized vortex ring.
Using our newly developed approach, this characteristic structure is clearly observed.
%
\begin{figure}[b!]
	\includegraphics [width=1\columnwidth]{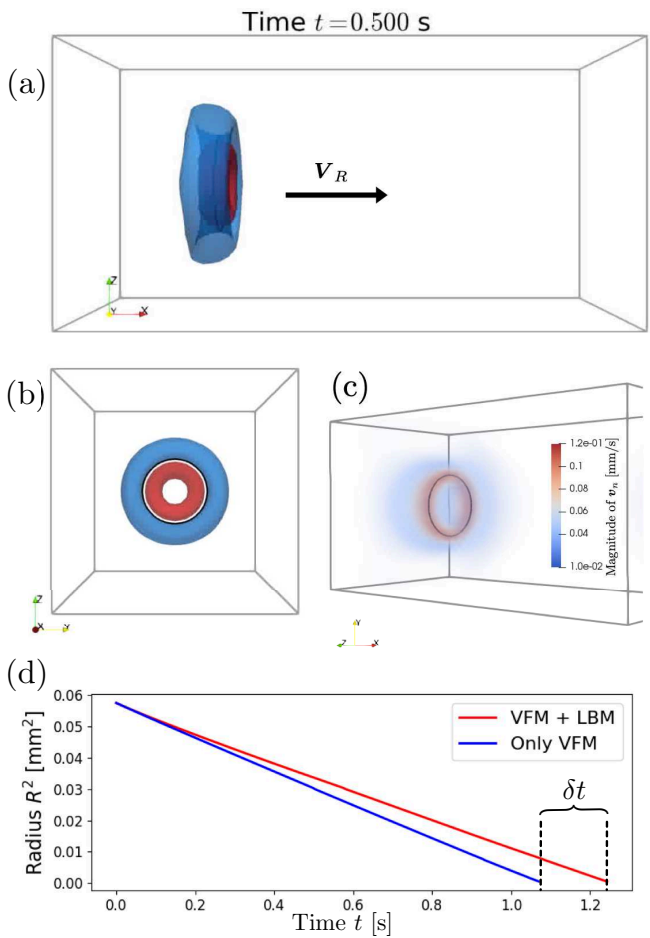}
	\caption{(a)--(b) Triple vortex structure at time $t = 0.5$ s. The black solid curve represents a quantized vortex ring. The red and blue surfaces represent vorticity isosurfaces in cylindrical polar coordinates. (c) Distribution of magnitude of normal fluid around quantized vortex. The color in the velocity is set such that the high-velocity region is red and thick and becomes blue and transparent for low velocities. (d) Comparison between coupled and conventional VFM simulations at $T = 1.9$ K. 
The lifetime of the vortex ring is elongated by $\delta t$.  
%
 }
	\label{fig: OneRing}
\end{figure}

\begin{figure*}[t!]
	\includegraphics [width=2\columnwidth]{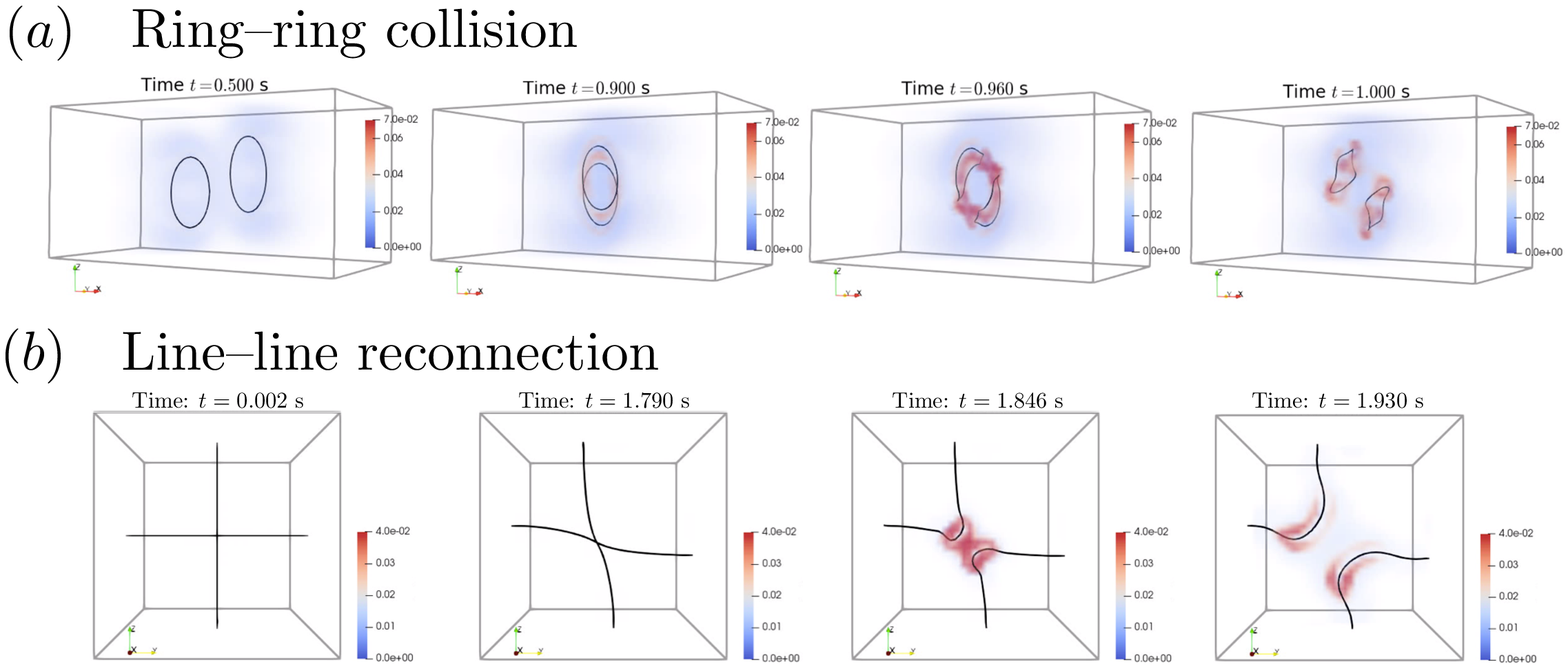}
	\caption{(a)--(b) Snapshots of vortex reconnection events and normal-fluid vorticity distribution: (a) Collision of two vortex rings and (b) vortex reconnection between two linear vortices. In each panel, the regions with low vorticity are set to be transparent such that the high-vorticity region (in red) can be easily observed.}
	\label{fig: Reconnection}
\end{figure*}
Figure \ref{fig: OneRing} displays a snapshot of the simulation at $t = 0.5$ s (the entire movie of the simulation can be found in \cite{Sup1}).
The initial radius of the quantized vortex ring is $R = 0.24$ mm, or in the lattice unit $R^\textrm{LB} = 7.68 \Delta x$.
The vortex ring travels with velocity $ \B{V}_R$, ``scattering'' normal-fluid component.
Considering a cylindrical polar coordinate system whose polar axis corresponds to the axis perpendicular to the vortex ring and passes through its center, the vorticity field, defined as $\B{\omega} = \nabla \times \B{v}$, can be translated from the Cartesian coordinates $(\omega_x,\omega_y,\omega_z)$ to the cylindrical coordinate $(\omega_\rho,\omega_\phi,\omega_{z^\prime})$.
Two isosurfaces (red and blue) are indicated in Fig. \ref{fig: OneRing} (a) and (b).
These correspond to the iso-vorticity surfaces with $\omega_\phi = -0.02/ \Delta x$ s$^{-1}$ (red) and $\omega_\phi = 0.02/ \Delta x$ s$^{-1}$ (blue).
Here, $\Delta x$ is the lattice separation measured in the physical units.
The triple vortex structure is stable as mentioned in Ref. \cite{Kivotides00}; they travel together maintaining certain volumes inside the isosurfaces.
However, as the radius of the quantized vortex shrinks owing to the mutual friction, the inner normal-fluid vortex ring is squeezed and eventually vanishes, whereas the outer one elongates in the propagating direction.
After the inner ring disappears, the quantized vortex ring travels for a small distance, dragging the normal-fluid component (see Fig. \ref{fig: OneRing} (c) and movie in \cite{Sup1}), and then vanishes.
%

The LIA allows us to estimate the time evolution of the vortex ring $R$ to be 
\begin{equation}
R = \sqrt{  R_0^2 - 2  \alpha \beta_\textrm{ind} t},
\end{equation}
where $\beta_\textrm{ind} \equiv \frac{\kappa}{4\pi}\ln \frac{R}{a}$, $a$ is the vortex core size and $R_0$ is the initial vortex ring radius.
Time evolutions of the square of the radius $R$ for the coupled simulation (VFM + LBM) and conventional VFM simulation at $T=1.9$ K are displayed in Fig. \ref{fig: OneRing} (d).
%
%
The fact that the slope of the coupled simulation is less steep indicates that the friction is smaller effectively because the vortex ring tends to drag the normal fluid around it. 
The difference $\delta t$ in the lifetime depends on the temperature $T$ of the system because of the temperature dependence of the mass densities $\rho_s$ and $\rho_n$ which is consistent with resent study by Galantucci \et{} \cite{Galantucci21}.

%
%
%

\subsection{Reconnections}\label{subsec: Reconnections}
Based on Eq. \eqref{eq: Biot--Savart} with support of the LIA, the velocity at which a quantized vortex travels is inversely proportional to its local radius of curvature.
This indicates that the motions/velocities of the vortices abruptly change during vortex reconnection events.
Figures \ref{fig: Reconnection} (a)--(b) display two typical types of reconnection events between vortices.
In each panel, the quantity $|\B{\omega}| \Delta x$ [mm/s] is plotted in the same manner as in Fig. \ref{fig: OneRing} (c).
%

While the radius of curvature is large, the vortex filaments move slowly and do not ``stir'' the normal fluid strongly.
As the two vortices approach each other, the vorticity of the normal fluid gradually accumulates in the region between them (see Fig. \ref{fig: Reconnection} (a) at $t = 0.9$ s).
When two vortices are within the distance of the resolution $\Delta \xi_\textrm{max}$, we algorithmically exchange their legs, which creates ``kinks'' on the filaments.
Because the kinks move with considerably greater local velocities, the normal fluid flow is also excited locally (see Fig. \ref{fig: Reconnection} (a) at $t = 0.96$ s and (b) at $t = 1.85$ s).
At finite temperatures, however, small structures on the vortex filaments are damped and the strongly excited region of the normal fluid diffuses and its magnitude decreases.
%

\begin{figure}[t!]
	\includegraphics [width=0.9\columnwidth]{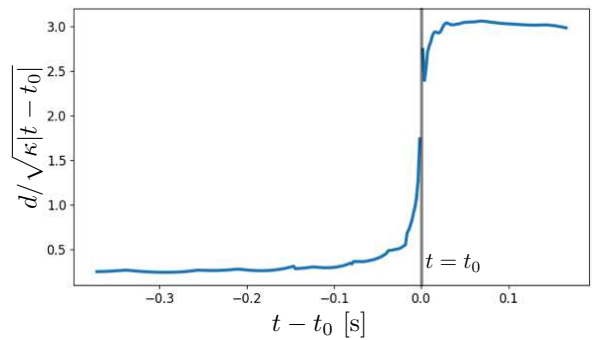}
	\caption{Dynamical scaling of vortices after reconnection. The result is obtained from the numerical simulation in \ref{fig: Reconnection}(b). The line-line reconnection occurs at $t = t_0$.}.
	\label{fig: Reconnection_A}
\end{figure}

We also confirm that the dynamical scaling behavior of the vortices after the vortex reconnection event is not altered by the presence of the normal fluid.
The only characteristic spatial and temporal scale of a quantized vortex is the circulation quantum $\kappa $ (dimensions of $L^2/T$).
Thus, the closest distance $d$ between the two vortices after reconnection can be written as 
\begin{equation}
d = A \sqrt{ \kappa ( t-t_0)},
\end{equation}
where $A$ is a dimensionless number of order unity and $t_0$ is the time at which reconnection occurs.
Figure \ref{fig: Reconnection_A} displays the value $A$ as a function of time, which agrees with the values known experimentally and numerically \cite{Fonda19,Minowa21}.
%

%

In the case of a highly dense vortex tangle, such as that discussed in the next section, these types of reconnection events occur at numerous locations in the system.
Thus, even when the normal-fluid flow as a whole is laminar, we expect certain non-negligible fluctuations in the flow that arise from the local ephemeral excitations of the normal-fluid velocity field.
A more detailed analysis of this effect and a comparison with previous studies are presented in the following section. 

\subsection{Vortex Tangle ($T$-1 regime)}\label{subsec: Vortex Tangle}
\begin{figure}[t!]
	\includegraphics [width=1\columnwidth]{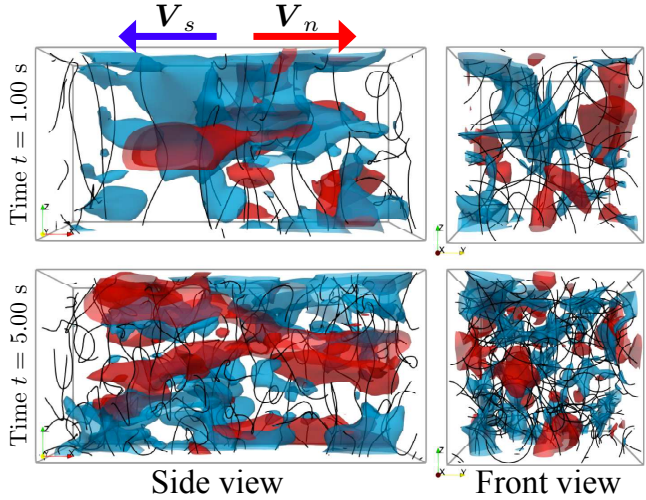}
	\caption{Development of vortex tangle and fluctuation in normal-fluid velocity profile at $t = 0, 1, 5$ s. The black curves represent vortex filaments; all walls are subject to the periodic boundary condition. Here, the average normal-fluid velocity $|\B{V}_n|$ is set to $2.5$ mm/s at the wall perpendicular to the counterflow direction. The blue and red surfaces are velocity fluctuation isosurfaces that correspond to $|\B{v}_n| = 0.9 \times|\B{V}_n|$ and $|\B{v}_n| = 1.1 \times|\B{V}_n|$, respectively. Note that our definition of isosurface is marginally different from that presented by Yui \et{} in Ref. \cite{Yui20}.}
	\label{fig: T1_2pt5}
\end{figure}

A recent numerical study by Yui \et{} \cite{Yui20} is considered in this section.
They performed a series of numerical calculations based on NS equations to simulate the coupled dynamics of superfluid and normal fluid, and to reveal the origin of anomalous fluctuations in normal fluid observed in the $T$-1 state \cite{Mastracci18}.
Their results, consistent with particle tracking velocimetry (PTV) measurements \cite{Mastracci19}, exhibited stronger normal-fluid fluctuations in the streamwise direction than those in the transverse direction.
The parameters in our simulation are adjusted to those in their study to ensure that the validity of our results with the newly developed numerical approach can be assessed quantitatively.
Except for the numerical approaches, there are two minor differences: one is the configuration of the initial seed vortices and the other is the boundary condition of walls perpendicular to the counterflow direction (along $x$-axis).
Because we are interested in the statistically steady vortex tangle, the first difference should be trivial, \ie{}, the steady state should not depend on the initial state.
The second difference could potentially be significant.
In this study, to create a steady thermal counterflow, we first perform an LBM simulation without vortices by manually setting the velocities along the $x$-axis at lattice sites near the walls perpendicular to $x$-axis.
This results in a reduction in the fluctuation in the normal-fluid flow profile near the walls, which can be observed in Fig. \ref{fig: T1_2pt5} near the walls perpendicular to the counterflow direction at $t = 5$ s.
%

Figure \ref{fig: T1_2pt5} illustrates the development of the statistically steady vortex tangle.
The boundary condition of the system is periodic at each wall.
As mentioned above, the thermal counterflow velocity profile is obtained by running an LBM simulation without vortices in advance.
The normal-fluid velocity $\B{V}_n$ due to heating along the channel ($x$-axis) is maintained by manually setting the velocities at the lattice points at $x = 0$ and $x=L_x$ to be $2.5$ mm/s.
In the coupled simulation, the normal-fluid velocity is constantly disturbed by the vortex filaments, which can be expressed as:
\begin{equation}
\B{v}_n(\B{x},t) = \B{V}_n + \delta \B{v}_n(\B{x},t). 
\end{equation}
The superfluid velocity $\B{V}_s$ induced by the heater at an arbitrary time $t$ is calculated by averaging $\B{v}_n$ and imposing the conservation of mass:
\begin{equation} \label{eq: mass cons}
\B{V}_s(x,t) = -\frac{\rho_n}{\rho_s} \frac{1}{A} \int\int \B{v}_n(\B{x},t) dy dz,
\end{equation}
where $A \equiv L_y L_z$ is the cross-sectional area of the channel.
Basically, the isosurfaces in the panels correspond to the surface where the magnitude of $\delta \B{v}_n$ is approximately $10$\% of that of $\B{V}_n$.
The blue surfaces are drawn where $\B{v}_n = 2.25$ mm/s; the red surfaces are drawn where $\B{v}_n = 2.75$ mm/s.
As the population of the vortex filaments increases, their spatial distribution is inhomogeneous. 
Because the filaments drag the normal fluid, $|\B{v}_n|$ tends to become smaller than $|\B{V}_n|$ in the region where the vortex line density is high.
Thus, the conservation of mass forces the region with a low vortex line density to satisfy $|\B{v}_n| > |\B{V}_n|$. 

\begin{figure}[t!]
	\includegraphics [width=1\columnwidth]{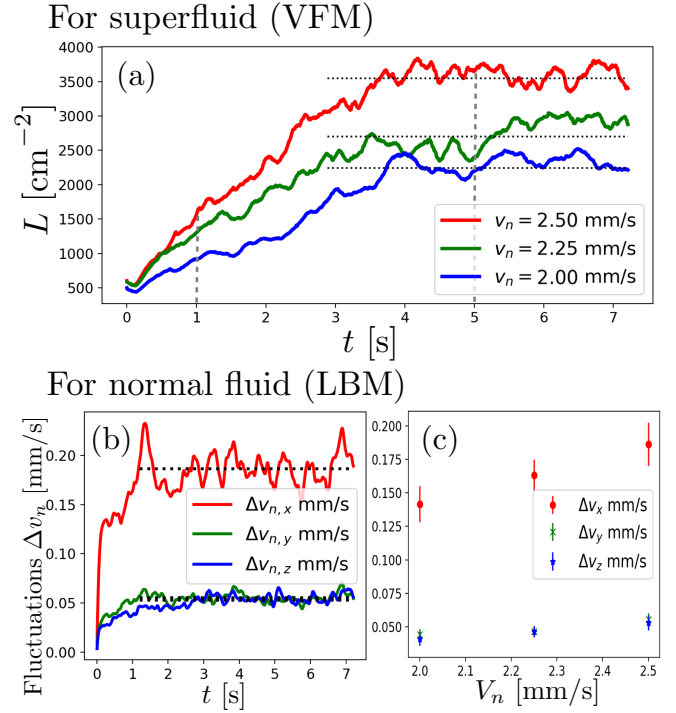}
	\caption{Quantitative assessment for the $T$-1 regime simulation. (a) Vortex line density $L$ as function of time $t$ for cases where $|\B{V}_n| =2.00$ mm/s, $|\B{V}_n| =2.25$ mm/s and $|\B{V}_n| =2.50$. The dashed lines are drawn at time $t = 1$ s and $5$ s. The corresponding snapshots for $|\B{V}_n| =2.5$ mm/s are displayed in Fig. \ref{fig: T1_2pt5}.  (b)--(c) Fluctuations in the normal-fluid velocity profile. The fluctuations observed here quantitatively agree with those obtained in the NS based simulations by Yui \et{} (see Ref \cite{Yui20} and Fig. 3 therein).}
	\label{fig: VLD}
\end{figure}

Now, we discuss the validity of our numerical approach quantitatively by comparing the  values of some physical quantities presented by Yui \et{} in Ref. \cite{Yui20}. 
Especially, the quantities we examine are the vortex line density $L$ (to check the validity of VFM) and the fluctuations in normal-fluid flow (to check the validity of LBM), and the results are summarized in Fig. \ref{fig: VLD} (a)--(c).
Because the vortex line density $L$ appears to saturate to a certain value depending on $|\B{V}_ n|$, the quantized vortex tangle developed under this setup is statistically steady.
Around $t = 4$ s for $|\B{V}_ n| = 2.5$ mm/s, the vortex line density $L$ saturates to and fluctuates around $L_\textrm{sat}\approx3500$ cm$^{-2}$ based on the $L$ vs. $t$ plot in Fig. \ref{fig: VLD} (a). 
Similarly, for $|\B{V}_ n| = 2.0$ mm/s, the vortex line density $L$ saturates to a certain value around $L_\textrm{sat}\approx2000$ cm$^{-2}$.
These values are consistent with the values obtained by Yui \et{} (see, for example, Fig. S. 2. in Ref. \cite{Yui20}).
Also, we verify that the square root of $L_\textrm{sat}$ is proportional to the relative velocity $V_{ns} \equiv |\B{V}_ n-\B{V}_ s|$ with the proportionality constant $\gamma = 142 \pm 17$, which is consistent with that obtained by Yui \et{} within the margin of error.

The anisotropy in the fluctuations in the streamwise and transverse directions, which was experimentally observed in Refs. \cite{Mastracci18, Mastracci19} and numerically obtained in Refs. \cite{Biferale19, Yui20}, is also confirmed in our numerical simulation as a stripe-like structure elongated in the streamwise direction (see Fig. 4 and video in Ref. \cite{Sup1}).
The spatial average of fluctuations $\Delta v_{n}$ in the normal-fluid flow profile is displayed in Fig. \ref{fig: VLD} (b)--(c) component-wise.  Although the examined parameter-range is smaller than the work by Yui \et{}, we can confirm a quantitative agreement between them.
%

\section{Discussion: Thermal Boundary Layer} \label{sec: DIS}
\begin{figure}[t!]
	\includegraphics [width=1\columnwidth]{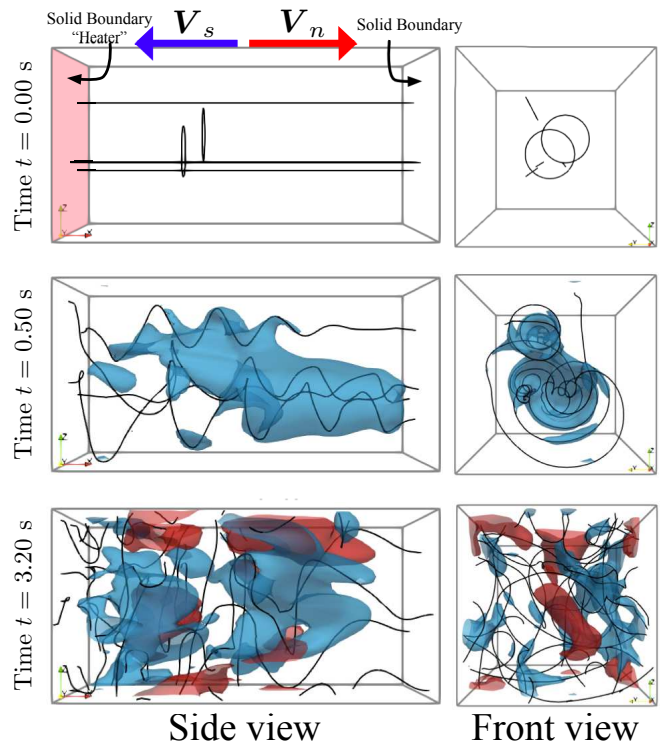}
	\caption{Snapshots of vortex tangle evolution at time $t = 0,0, 0.5, 3.2$ s. The coloring scheme in the panels is same as in Fig. \ref{fig: T1_2pt5}.}
	\label{fig: Wall_density_anim}
\end{figure}

%
The LBM is one of the most appreciated methods for CFD and has been used for various purposes.
In the LBM framework, we simulate a fluid flow with local operations \ie{}, a series of convections and collisions among the fluid particles.
Because the performance of the simulation is dependent on the local operations, the spatial and temporal resolutions of the simulation are critical to the reproduction of a realistic flow with the LBM, which could be raised as a major disadvantage of adopting this approach.
However, there are also advantages, one of which is that the LBM allows complex boundary conditions that may not be reasonably justified in NS simulations.
An example of such a situation is discussed below where we consider a closed channel geometry, as displayed in Fig. \ref{fig: Wall_density_anim}.
Similar to that displayed in Fig. \ref{fig: T1_2pt5}, there is a thermal counterflow, $\B{V}_s$ and $\B{V}_n$, excited along the channel, and isosurfaces are drawn to indicate the surfaces on which  $|\B{v}_n| = 0.9 \times|\B{V}_n|$ (blue) and $|\B{v}_n| = 1.1 \times|\B{V}_n|$ (red).
 However, in this case, the walls perpendicular to the streaming direction are subject to the solid boundary condition, \ie, the mass conservation in Eq. \eqref{eq: mass cons} is strictly enforced at the boundaries, whereas the other walls are periodic.
 At a glance, the difference between two cases (closed channel and open/periodic channel) may seem to be minor; however, since the wall of the closed channel behaves as a heater/source of normal-fluid component, its treatment is not trivial in NS simulations.  
 On the other hand, within the framework of LBM, a steady normal-fluid efflux from the wall can be simply implemented by adjusting the population of (mesoscopic) fluid particles near the wall.
%

%
This setup is inspired by the recent experimental and numerical study conducted by Varga and Skrbek \cite{Varga19}, in which they discuss the ``boundary layer'', where the temperature gradient is increased, is formed near a heater.
They claim that the origin of the rise in the temperature gradient can be partly ascribed to the inhomogeneous vortex line density near the heater induced by the inhomogeneity in the heat flux.
 At a steady state, the relationship between the temperature gradient $\nabla T$ and vortex line density $L$ can be approximately obtained as \cite{Varga19}
\begin{equation} \label{eq: grad T}
\nabla T = -\frac{\alpha}{\sigma}\kappa L (\B{v}_n - \B{v}_s).
\end{equation}
In their study, the inhomogeneity in the counterflow profile is prescribed and static; we make the counterflow profile inhomogeneous automatically by adopting the LBM for normal-fluid time evolution.
In the simulation results presented below, we consider only the simplest heater geometry, \ie{}, homogeneously heated/cooled walls at the ends of the channel.
More involved heater geometries such as those considered in Refs. \cite{Varga19,Rickinson20,Inui20_1} are of interest for future works.
%

%
\begin{figure}[t!]
	\includegraphics [width=1\columnwidth]{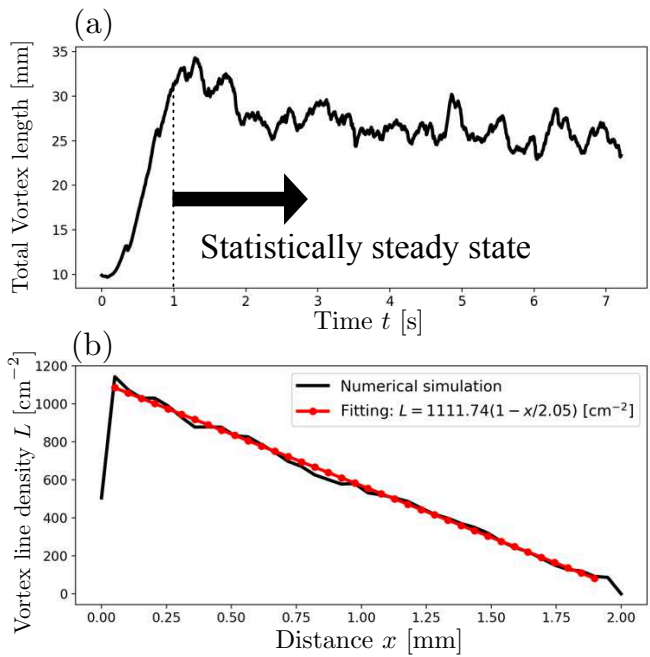}
	\caption{(a) Total vortex line length as function of time $t$. Although the vortex tangle is not homogeneous, it achieves a statistically steady state and the total length appears to saturate at approximately $t=1$ s. (b) Vortex line density $L$ as a function of distance $x$ from solid boundary. The value of $L$ is obtained by averaging all vortex configurations after $t=1$ s.}
	\label{fig: Wall_density}
\end{figure}

Figure \ref{fig: Wall_density_anim} displays a time evolution of seed vortices under a counterflow originating from the solid walls.
Initially, four vortex lines are bound to the solid walls, and two smaller rings are placed to provide a ``kick'' to the linear vortices (see Fig. \ref{fig: Wall_density_anim} at time $t = 0$ s ).
Under the thermal counterflow with $|\B{V}_n| = 2.5$ mm/s at $T = 1.9$ K, the linear vortex line becomes unstable and a helical Kelvin wave is excited.
The helical excitations are amplified as they travel in the downstream direction of the superfluid component, which can be observed as helical cone-like structures at $t = 0.5$ s in Fig. \ref{fig: Wall_density_anim}.
The helically deformed vortex filaments disturb the normal-fluid flow behind it and a large portion of the helical cone-like structure is enclosed by blue isosurfaces ($|\B{v}_n| = 0.9 \times|\B{V}_n|$).
The conservation of mass, however, induces fluid flow in the other direction, where the vortex line density is relatively low.
Then, the region enclosed by the red surfaces ($|\B{v}_n| = 1.1 \times|\B{V}_n|$) starts to develop and the quantized vortex tangle eventually achieves a statistically steady state.
In the steady state, unlike a vortex tangle developed under a counterflow in a periodic boundary along the streaming direction, the vortex line density is no longer spatially homogeneous and the invariance under translation along the streaming direction is broken.
Figure \ref{fig: Wall_density} (a) plots the total vortex filament length.
Even though the fluctuation remains relatively large, the total length appears to saturate to a certain value at approximately $t = 1$ s.
The vortex line density $L$ along the streaming direction ($x$-axis) is displayed in Fig. \ref{fig: Wall_density} (b) as a function of distance $x$ measured from the wall on the left end.
The density $L$ is obtained by averaging all vortex configurations from $t=1.0$ s to $t = 7.2$ s.
%

If we replace $\B{v}_n$ and $\B{v}_s$ in Eq. \eqref{eq: grad T} with the averaged quantities, $\B{V}_n$ and $\B{V}_s$, and consider $\sigma$ constant, we can easily solve the ordinary differential equation for $T$ as the density $L(x)$ is approximately linear in $x$.
With $L(x) = L_0 (1-x/x_0)$, the solution to Eq. \eqref{eq: grad T} is
\begin{equation}\label{eq: Temp profile}
T(x) = T_0  -  \delta T  \left( 1 - \frac{x}{2x_0}  \right)x,
\end{equation}
where $\delta T \equiv \frac{\alpha}{\sigma}\kappa L_0 V_{ns}$ and $V_{ns} \equiv | \B{V}_n - \B{V}_s |$.
Using the values for $L_0$ and $x_0$ estimated by fitting the numerical results, as indicated in Fig. \ref{fig: Wall_density} (b), the temperature difference $\delta T \approx  1.40 \times 10^{-3}$ mK/m, is less than the values obtained by Varga \et{} by approximately two to three orders of magnitude.
The major reason for this discrepancy seems to arise from the fact that the local vortex line density $L(x)$ does not reach highest attainable value $L_ \textrm{sat} \approx 3500$ mm$^{-2}$ (when the channel is periodic ) because the ``entrance length'' of the channel is not sufficiently long to achieve the value.
A more detailed discussion on the entrance length problem will be presented in a forthcoming paper currently in preparation.
It is important to note here that this LBM-based numerical approach allows the estimation and implementation of spatial temperature dependencies on physical quantities such as friction coefficients, $\alpha$ and $\alpha^\prime$.  
This could lead to remarkable progress in the superfluid community as the spatially inhomogeneous temperature profile is critical for understanding the formation of vortex tangles around a localized heater.
 In Ref. \cite{Sergeev19}, Sergeev \et{} suggest that if one allows the mutual friction constants to vary spatially, a statistically steady vortex tangle could be found around a cylindrically symmetric heater based on the HVBK simulation.
Using the temperature profile they obtained, Rickinson \et{} \cite{Rickinson20} performed VFM simulations and demonstrated that a statistically steady turbulence state exists.
This is a remarkable achievement in the study of inhomogeneous vortex tangles; however, their numerical simulation relies on a result based on a different framework.
The LBM-based approach, on the other hand, allows us to perform a stand-alone temperature dependent simulation.

\section{Summary} \label{sec: SUM}
%
%
We developed a computational method for two-fluid coupled dynamics of superfluid \He{} at finite temperature and examined its performance by comparing it with numerical simulations performed with different numerical approaches \cite{Kivotides00,Yui20}.
The proposed computational approach for solving the normal-fluid flow is based on the LBM, which is widely adopted in CFD, yet rarely applied in the study of superfluid \He{} except for the recent study by Bertolaccini \et{} \cite{Bertolaccini17}.
In their study, they adopted a two-dimensional LBM simulation for both normal fluids and superfluids.
Conversely, the proposed computational approach is as follows: superfluid flow is obtained using the VFM and normal-fluid flow computation is performed by a three-dimensional LBM.
%

%
When a quantized vortex ring passes through a normal fluid at rest, two normal-fluid vortical structures are formed immediately inside and outside the vortex ring, as reported and referred to as ``triple vortex ring structure'' in Ref. \cite{Kivotides00}.
As the quantized vortex ring shrinks owing to the mutual friction, its propagation velocity increases.
Therefore, the outer normal-fluid vortex is elongated in the propagation direction, whereas the inner vortex eventually vanishes.
We also demonstrated the normal-fluid flow immediately before and after a vortex-vortex reconnection event.
As can be observed from the previous example, the smaller the local curvature radius $R$ of a quantized vortex becomes, the more strongly the normal-fluid flow is disturbed as the superfluid flow induced by the vortex filament is proportional to $1/R$.
In the snapshots displayed in Fig. \ref{fig: Reconnection}, the ephemeral excitations in normal-fluid flow caused by typical reconnection events are summarized. 
In a vortex tangle, as discussed in Sec. \ref{subsec: Vortex Tangle} and Sec. \ref{sec: DIS}, such ephemeral excitations are continuously created and the inhomogeneity in fluctuations remains persistently in the normal-fluid flow profile, which qualitatively agrees with phenomena reported in Refs. \cite{Mastracci18, Mastracci19, Biferale19, Yui20}.
%

%
In the LBM approach, the boundaries can be flexibly considered, unlike numerical approaches that directly solve the NS equation.
Thus, we performed a counterflow simulation with a solid boundary in the streaming direction.
Even under such conditions, a statistically steady vortex tangle was maintained, although the vortex line density was highly inhomogeneous. 
Inhomogeneity in the tangle is essential for explaining the boundary layer near a heater \cite{Varga19}; hence, we can conclude that the proposed computational approach is well suited to numerical simulation with this type of complex boundary condition.
%

\begin{acknowledgements}
M.T. acknowledges the support from JSPS KAKENHI (Grant No. JP20H01855). S.I. was supported by a Grant-in-Aid for JSPS Research Fellows (Grant No. JP20J23131).
\end{acknowledgements}


\bibliography{LBMC}

\end{document}